\documentclass[12pt]{iopart}

\usepackage{iopams}

\usepackage{amssymb,bm}
\usepackage{amsthm} 

\usepackage{graphicx}
\usepackage{psfrag}
\usepackage{epsfig}

\def\unity{\mathbb{I}}
\def\ra{\rangle}
\def\la{\langle}

\def\be{\begin{equation}}
\def\ee{\end{equation}}

\usepackage[T1]{fontenc}
\usepackage{mathptmx}

\usepackage[dvipsnames]{color}
\usepackage[colorlinks=true,citecolor=myblue,linkcolor=myblue,urlcolor=myblue]{hyperref}
\definecolor{myblue}{named}{MidnightBlue}
\definecolor{gray}{named}{Gray}

\newcommand{\bra}[1]{\left\langle{#1}\right\vert}
\newcommand{\ket}[1]{\left\vert{#1}\right\rangle}

\theoremstyle{definition} 
\newtheorem{definition}{Definition}

\theoremstyle{definition} 
\newtheorem{criterion}{Criterion}

\begin{document}

\title[What is a quantum computer, and how do we build one?]{What is a quantum computer, and how do we build one?}  
\author{Carlos A. P\'erez-Delgado\footnote{c.perez@sheffield.ac.uk} and Pieter Kok\footnote{p.kok@sheffield.ac.uk}}
\address{Department of Physics and Astronomy, University of Sheffield, Hicks building, Hounsfield road, Sheffield, S3 7RH, United Kingdom}

\begin{abstract}
\noindent 
The DiVincenzo criteria for implementing a quantum computer have been seminal in focussing both experimental and theoretical research in quantum information processing. 
These criteria were formulated specifically for the circuit model of quantum computing. 
However, several new models for quantum computing (paradigms) have been proposed that do not seem to fit the criteria well.
The question is therefore what are the general criteria for implementing quantum computers.
To this end, a formal operational definition of a quantum computer is introduced. 
It is then shown that according to this definition a device is a quantum computer if it obeys the following criteria: 
Any quantum computer must consist of a quantum memory, with additional structure that (1) facilitates a controlled quantum evolution of the quantum memory; (2) includes a method for information theoretic cooling of the memory; and (3) provides a readout mechanism for subsets of the quantum memory.
The criteria are met when the device is scalable and operates fault-tolerantly. 
We discuss various existing quantum computing paradigms, and how they fit within this framework.
Finally, we present a decision tree for selecting an avenue towards building a quantum computer.
This is intended to help experimentalists determine the most natural paradigm given a particular physical implementation.
\end{abstract}


\submitto{New Journal of Physics}

\date{\today}
\maketitle


\section{Introduction}

\noindent
One of the main focal points of modern physics is the construction of a full-scale quantum computer \cite{deutsch85,deutsch89}, which holds the promise of vastly increased computational power in simulating quantum systems. In turn, this may lead to fundamentally new quantum technologies \cite{feynman82,lloyd96}.
There is also mounting evidence that these devices can solve standard mathematical-computational problems more efficiently than classical computers.
Improvements range from quadratic speed-up in general-purpose algorithms such as search \cite{BBBGLS00,Grover96,BBBGL98,BBBGL99,CGW00,Grover98}, to exponential (over best known classical counterparts) in specialized algorithms, such as the hidden subgroup problem \cite{shor94,CEHM98,CEMM98}.

The common goal of creating a quantum computer has acted as a focus for research in experimental and theoretical physics, 
as well as in computer science and mathematics.
In particular, this goal has pushed forward the search for better ways to control different types of quantum systems, like cold and hot ions  \cite{cirac95,molmer99}, Cavity QED  \cite{pellizari95}, neutral atoms \cite{briegel00},
liquid and solid state NMR \cite{cory97,gerschenfeld97,braunstein99,yamaguchi99,vandersypen04}, silicon-based nuclear spins \cite{kane98}, electrons floating on Helium \cite{Platzman99}, quantum dots \cite{loss98}, Cooper pairs \cite{averin98}, Josephson junctions \cite{mooij99,schnirman97,makhlin99,makhlin01}, and linear optical systems  \cite{knill01,kok07} among others.
Important new results in the field, both theoretical and experimental, continue to drive progress in quantum information processing. 
As a consequence, our ability to control quantum systems  has improved dramatically over the past fifteen years, and we understand much better many aspects of classical and quantum computing, as well as broader aspects of fundamental physics.

An important part in focussing the research in quantum information processing has been the set of criteria for creating a quantum computer, pioneered by Deutsch \cite{deutsch85}, and  expanded and formalized by DiVincenzo \cite{divincenzo96,divincenzo00,divincenzo01}. 
They have inspired new experimental and theoretical research in quantum control and quantum information processing.
The criteria, now known as {\em DiVincenzo's Criteria}, apply explicitly to the circuit model of quantum computation. 
According to the criteria, any quantum computer must facilitate the following:
\begin{enumerate}\itemsep=0pt
 \item A scalable physical system with well-characterized qubits;
 \item the ability to initialize the state of the qubits to a simple fiducial state, such as $\ket{000\ldots}$;
 \item long relevant decoherence times, much longer than the gate operation time;
 \item a universal set of quantum gates;
 \item a qubit-specific measurement capability.
\end{enumerate}
These five criteria were originally formulated by DiVincenzo in 1996 \cite{divincenzo96}. Subsequently, DiVincenzo formulated two more criteria for quantum communication \cite{divincenzo00}:
\begin{enumerate}
 \item[(vi)] the ability to interconvert stationary and flying qubits;
 \item[(vii)] the ability to faithfully transmit flying qubits between specified locations.
\end{enumerate}

Since the formulation of the criteria, new ways of making quantum computers have been invented that do not always seem to fit the criteria very well.
Today, experimentalists can choose between various paradigms for quantum computation, such as adiabatic quantum computing, globally controlled quantum computing, and the one-way model, all of them with their various strengths and weaknesses.
Each paradigm requires a completely different approach, yet all attempt to reach the same end goal, namely to construct a quantum computer.
While all models are computationally equivalent, their differences allow for different intuitions, and practical advantages.
This allows for greater freedom in the laboratory. 
For example, where one was once required to control individual qubits, today we know that global control suffices in certain instances.
The wealth of paradigms also enriches our theoretical knowledge, and increases our chances for finding new algorithms. 
For example, universal blind quantum computation \cite{broadbent08} was developed using intuition gained from the one-way model for quantum computing. 
DiVincenzo himself wrote about the implications of these new paradigms for the criteria in 2001, and relished how heresies to the ``dogmatic'' criteria were arising in the field of quantum computation \cite{divincenzo01}.

We expect that there must be a form of the criteria that does not make any assumptions about the particular implementation of the quantum computer.
In this paper, we formulate such general criteria for constructing a quantum computer, and we identify the metric that determines whether the criteria are met in terms of fault tolerance and scalability.
To achieve this, we must construct a new {\em operational} definition of a quantum computer, and the algorithms that run on them.
Finally, we construct a decision tree to help select the most promising paradigm given a particular physical implementation. 
This will provide the experimentalist with a map of the key theoretical results they need in order to make their laboratory setup into a full-scale quantum computer. 
While the criteria are designed to be independent of the paradigms, the decision tree must be updated whenever new paradigms are developed.

\section{What is a Quantum Computer?}

\noindent
Before we can discuss the new criteria for quantum computation, we have to define exactly what we mean by the term {\em quantum computer}.
Although most readers will have an intuitive concept, a formal definition of the term is quite elusive. 
In the literature, there are broadly four types of definitions, and we will argue that they generally fall short of what we seek in a useful definition of a quantum computer.
First, a quantum computer can be defined as a representation of a quantum Turing machine, as proposed by Deutsch in 1985 \cite{deutsch85}. While satisfactory from a formal computer science perspective, this is not the most useful formulation when one is concerned primarily with the implementation of a quantum computer. Second, many texts use an implicit definition of a quantum computer. For example, Mermin \cite{mermin07} writes that ``a quantum computer is one whose operation exploits certain very special transformations of its internal state'', and the subject of the book is these special transformations. In Nielsen and Chuang \cite{nielsen00}, a definition of a quantum computer is never given, and the reader instead develops an intuition for the meaning of the word ``quantum computer'' over the course of the material. The implicit definition is a perfectly good pedagogical approach, but it is not the clear, brief statement we need to derive the criteria. Third, quantum computers are sometimes defined as devices that can outperform classical computers. The trouble with these types of definitions is that they depend on the classification of computational problems, and there are a number of important open questions about this classification (for example whether $\mathsf{P} = \mathsf{NP}$). Although unlikely, it may well be that classical computers are just as powerful as quantum computers, and the definition ceases to have meaning. We therefore require a definition that does not depend on the classification of problems in complexity theory. Finally, there are constructive definitions, which state that quantum computers are made of quantum bits, use entanglement, etc. The trouble with such definitions is that they tend to be quite specific about the implementation. For example, a definition in terms of qubits seems to exclude the possibility of creating a quantum computer using continuous variable quantum systems.   Instead, we want a definition that does not presuppose anything about the building blocks of the quantum computer, does not depend on the classification of problems in complexity theory, and is independent of our interpretation of quantum mechanics (two philosophers with radically different interpretations of quantum mechanics should still be able to agree whether a device can be classified as a quantum computer). Finally, the definition must not make any reference to the paradigm that is used to perform the computation in any fundamental way. For example, the end user of a quantum computer will generally not be interested whether the device uses the adiabatic, measurement-based, or some other as yet unknown form of quantum computing, unless that makes a difference in the performance of the device. In other words, we need an {\em operational} definition of a quantum computer. We will give a formal definition of a quantum computer below, after a brief discussion of the intuitive background of this definition. 

Broadly speaking, we define a quantum computer as a device that can run a quantum algorithm efficiently, where a quantum algorithm is a classical bit string that encodes a series of quantum operations (typically quantum gates).
The quantum computer should be able to take this string as input and produce as output another bit string. 
The probability distribution of the output should be consistent with the predictions of quantum theory. 
Finally, the time it takes the computer to produce the output should be in agreement with the difficulty of the algorithm, \emph{e.g.} an exponential-time algorithm can take the quantum computer an exponentially long time to compute, but a polynomial-time algorithm should be computed in polynomial time. To see how a classical computer is believed to fail this criteria, consider Shor's factoring algorithm: The number of steps in the algorithm scales polynomially with the size of the input, but the actual classical implementation will scale exponentially. 

There are three important advantages to this definition.
First, it makes no reference to how the quantum computer works, which means that the definition does not have to be updated when new methods of implementation are invented.
Second, the definition does not refer to any specific computational problems, like factoring, as a differentiator from a classical computer.
Instead, the definition calls for the ability to compute \emph{any} quantum algorithm, in a efficient enough manner.
And third, the definition does not make any assumptions about either the theory of  computation or the nature of physical reality.
This means that the definition will still be valid when our knowledge of the relationship between classical and quantum computing becomes more complete, and when physical theories that supersede quantum theory are developed.

To formulate our definition in a precise mathematical way, let $\smash{s^{(n)}_{\rm in}}$ be a string of classical symbols, and let the program $P$ of size $r$ be a symbolic representation of an algorithm (for more details see Sec.\ \ref{sec:algorithms}).
\begin{definition}\label{def:idealQC}
An {\em ideal quantum computer} is a hypothetical device that accepts as input a classical bit string $\smash{s^{(n)}_{\rm in}}$, and a quantum program $P$ with size $r$, acting on a Hilbert space $\mathcal{H}_n$ of dimension $2^n$. For any given program $P$ the quantum computer produces the classical output bit string $\smash{s^{(m)}_{\rm out}}$ with probability
 \begin{equation*}
  p_P (s^{(m)}_{\rm out}|s^{(n)}_{\rm in}) =  \la{s^{(n)}_{\rm in}} | U_P^{\dagger} \left( \unity_{n-m} \otimes |{s^{(m)}_{\rm out}}\ra\la{s^{(m)}_{\rm out}}| \right) U_P |{s^{(n)}_{\rm in}}\ra .
 \end{equation*} 
The total amount of resources used by the device scales polynomially in $r$.
\end{definition}
The operator $\unity_{n-m}$ is shorthand for the identity on the subspace of $\mathcal{H}_n$ that is the orthocomplement of the subspace spanned by $\{\ket{\smash{s^{(m)}_{\rm out}}}\}$.
It is worthwhile emphasizing that the quantum computer uses resources that scale polynomially in the size of the program $r$, as opposed to the number of bits $n$ in the input to the program.
In other words, a quantum computer can very well implement an exponential-time quantum algorithm.
The requirement states that if for example the algorithm itself is polynomial-time, then the quantum computer must also run in polynomial time.
We do not impose the stronger restriction that the quantum computer uses resources that are \emph{linear} in $r$.
There are a number of reasons why this is reasonable.
First, there are various different quantum computer paradigms, as we discuss elsewhere in this paper, and while all of them are equivalent with regards to which computation they can perform efficiently, some computations can be slightly more efficient on one platform than on another. 
For example, calculating the ground state of a BCS Hamiltonian on a traditional NMR quantum computer may be quadratically slower than doing the same calculation on a qubus quantum computer \cite{brown09}.
Second, as we have discussed earlier, there are various different languages in which quantum programs can written.
It may be necessary to translate or \emph{compile} the program from the language it is written to the \emph{`native'} language of the quantum computer; and this translation may take up to polynomial time in the size of the program.
Third, there are the issues of scalability and fault tolerance, to which we will return in Sec.\ \ref{sec:ftnsc}.
Because of overheads due to error-correction and fault-tolerance, it is possible that the amount of resources needed by  a quantum computer to solve larger problems scales super-linearly. 

Why is the quantum computer in Definition \ref{def:idealQC} a hypothetical device?
Suppose that instead we defined a real quantum computer according to Definition \ref{def:idealQC}.
The quantum computer should then be able to accept an input of any size, and compute arbitrary quantum programs on this input.
On the other hand, any real quantum computer has a well-defined finite size, in terms of the number of logical input bits it can operate on. 
Furthermore, it needs to potentially create entanglement across as many  subsystems as it has input bits.
We can therefore always construct a problem that  is too large for our (possibly very large) quantum computer, and such a device would fail according to Definition \ref{def:idealQC}. 
Nevertheless, any reasonable definition of a real quantum computer should admit this finite device as a true quantum computer. 

This is an important difference between classical and quantum computers.
Any classical computer that is large enough to be universal will be universal in the strictest sense, and no restrictions are placed on the efficiency of the classical computation.
It accepts an input string of any size and can compute an algorithm of any size, given that it is provided with sufficient `work space'.
We could introduce the notion of a `quantum work space', but that would be cheating. 
It is a profound  and important fact that a quantum computer cannot store partial results while it works on other parts of a computation.
Every part of a computation can potentially be entangled to every other part, which is central to the speedup that quantum computers can achieve over classical computers \cite{vidal03,jozsa03}.
Moreover, due to the monogamy of entanglement, stored partial results cannot interact with anything but the quantum computer itself.
The quantum work space must therefore be considered an integral part of the quantum computer. 

The quantum work space has traditionally been described as a set of qubits.
However, in recent years it has been shown that it is not necessary to restrict ourselves to the qubit model. 
In particular, an important sub-discipline of quantum information theory involves the use of continuous variable quantum information carriers, or {\em qunats}.
This therefore leads us to consider a formal definition of the quantum work space, which from now on we call a quantum memory:
\begin{definition}\label{def:qumemory}
 A {\em quantum memory} of size $k=\log_2 d$ is a physical system that can represent  any (computable) quantum state in a Hilbert space of dimension $d$.
\end{definition}

Even though the size of a quantum memory is measured in qubits, the definition does not specify what type of information carriers should be used.
However, $d$ must be finite, which means that even if the quantum memory is based on continuous variables, the effective Hilbert space must be finite.
Consequently, when continuous variables are the information carriers, the logical encoding must either be in terms of a qudit, or must take into account the finite precision that is inherent in continuous variables.
Finally, a subtle but important point:
According to Definition \ref{def:qumemory}, a quantum memory must be able to represent (that is, store) any computable quantum state, but we do \emph{not} require that it actually is in such a state.
That would be too strong a requirement. Just about any current experimental implementation  of a qubit is a good example of this. For instance, in NMR quantum computation the nuclear spin of an atom is used to represent a qubit. This is natural when using a spin-$1/2$ nucleus; but one can also use, say, spin-$3/2$ nuclei, and simply use a sub-space of the physical Hilbert space, as the `computational' space. A more extreme example would be optical lattices, or quantum dots, where experimentalists use two energy eigenstates of their respective systems to represent qubits. It does not matter that the energy eigenstates of an atom trapped in an optical lattice is infinite-dimensional; it can still represent a simple two-dimensional system. Finally, a more subtle example is a device that is not actually quantum mechanical itself, but can simulate any quantum mechanical interaction. This too is a perfectly viable quantum memory.

Let us now return to the definition of a quantum computer.
Another sense in which Definition \ref{def:idealQC} must describe a hypothetical device is that it must produce the probability distribution $\smash{p_P (s^{(m)}_{\rm out}|s^{(n)}_{\rm in})}$ exactly. 
However, any real device will unavoidably have errors, and can never produce $\smash{p_P (s^{(m)}_{\rm out}|s^{(n)}_{\rm in})}$ exactly.
Of course, from a practical point of view, this level of precision is quite unnecessary.
It is perfectly acceptable that the quantum computer creates $\smash{s^{(m)}_{\rm out}}$ with probability $\smash{p'_P (s^{(m)}_{\rm out}|s^{(n)}_{\rm in})}$, where $p_P$ and $p_P'$ are close to each other with respect to some metric (e.g., the fidelity or the statistical distance).
The difference between the two distributions will manifest itself as the occurrence of wrong answers in the quantum computer. 
If the problem to be solved is such that we can verify the correctness of the answer efficiently (such as factoring), then it is sufficient to run the quantum computer repeatedly.
As long as the probability of getting a wrong answer is small enough, we can repeat the computation until we obtain the right answer.
However, one might be interested in tackling problems that are not efficiently verifiable.
In addition, it is desirable to be able to lower bound the success probability of running the quantum computer efficiently. 
In such cases, we require more sophisticated methods of quality control.
One such method is  circuit self-testing \cite{magniez06}. Taking these practical considerations into account, we arrive at the following definition:
\begin{definition}\label{def:kqubitQC}
A {\em $k$-bit quantum computer} is a physical device that accepts as input a classical bit string $\smash{s^{(n)}_{\rm in}}$, with $n \leq k$, and a quantum program $P$ of size $r$, acting on a Hilbert space of size $2^n$.
For any given program $P$ the quantum computer  produces the classical output bit string $\smash{s^{(m)}_{\rm out}}$ with probability
 \begin{equation}\nonumber
  p_P (s^{(m)}_{\rm out}|s^{(n)}_{\rm in}) =  \la{s^{(n)}_{\rm in}} | U_P^{\dagger} \left( \unity_{n-m} \otimes |{s^{(m)}_{\rm out}}\ra\la{s^{(m)}_{\rm out}}| \right) U_P |{s^{(n)}_{\rm in}}\ra ,
 \end{equation} 
with sufficiently high fidelity. 
The amount of resources used by the device scales polynomially in $r$.
\end{definition}
This definition captures the notion of a quantum computer as a real, finite physical device.
In particular, we now allow our device to fail some of the time.
We impose the weaker condition that it reproduces the probability distribution with \emph{sufficiently} high fidelity.
Though we do not impose a particular number as the lower bound on the success probability, a standard choice would be $2/3$.
It is not \emph{a priori} clear that proving a lower bound on the reliability of a quantum device is easy, or even feasible. 
Practically, however, there are procedures that should be able to prove the reliability of some types of quantum computer architectures \cite{magniez06}.

The drawback of Definition \ref{def:kqubitQC} over Definition \ref{def:idealQC} is that the mathematical notion of scalability is lost, since the device is strictly finite.
To retrieve scalability, we must refer to Definition \ref{def:idealQC}.
That is, we can claim scalability of our quantum computer architecture when, given an input  string of any size $n$ and a quantum program, we can in principle construct a physical $k$-bit quantum computer with $k \geq n$ to run the computation.

Thus far, we have avoided any discussion on how the measurement statistics $\smash{p_P (s^{(m)}_{\rm out}|s^{(n)}_{\rm in})}$ are obtained. 
Definitions \ref{def:idealQC} and \ref{def:kqubitQC} are entirely operational, which means that if a device \emph{acts} like a quantum computer in producing the probability distribution, then by our definition it \emph{is} a quantum computer. 
Importantly, we have completely avoided any assumptions about the representation of the data in the quantum computer. 
Even though the input and output are always considered classical bit strings, the physical representation in the quantum computer can be in terms of qubits, qudits, or qunats.
Even if the physical representation is in terms of qubits, the logical qubits will in general not map directly to the physical qubits, for example due to levels of error correction. 
We have also sidestepped any consequences of possible efficient classical simulability of quantum processes. 
Finally, the definition is independent of interpretations of quantum mechanics.

\section{Defining Quantum Algorithms}\label{sec:algorithms}

\noindent
Our definition of a quantum computer is based on the notion of efficient computability. In other words, an efficient quantum program should be run efficiently by a quantum computer, while an inefficient program may run equally inefficiently.
The purpose of this section is to formalise this notion. However, for many readers the idea of a program $P$ of size $r$  will be sufficiently intuitive, and these readers can skip this section if they desire.

We formalise the notion of quantum program by means of an inductive definition: we begin by defining basic building-blocks and then describe how these building blocks can be combined to build more complex objects. A consequence of this is that we have to choose a particular way to \emph{construct} quantum algorithms.  We have chosen the language and building blocks of quantum circuits, since they form a well-known and intuitive approach. While in many ways it is desirable to give a general definition of quantum algorithms that does not refer to any particular  implementation, the definition presented here allows us to achieve the goals set in this paper properly and easily. An inductive definition allows us to define a formal size function in a straightforward manner. That is, not only is the size of a quantum program always well-defined, it is also easily characterised. In our case, it is simply the sum of the cost of each individual gate.

For simplicity, and without loss of generality, we will restrict ourselves predominantly to the binary alphabet $\{ 0, 1 \}$. 
Any higher-dimensional information carrier can be described in terms of multiple bits, and even continuous variables can be treated this way, as long as we keep in mind that they have an intrinsic finite precision.
Consequently, we consider mostly binary finite classical bit strings $s^{(n)} \in \{0, 1 \}^n$. Quantum bit strings will be denoted similarly by $\ket{\smash{s^{(n)}}} \in \{ \ket{0}, \ket{1} \}^{\otimes n}$.

We have defined quantum computers as accepting three pieces of input: a bit string representing the initial state; a bit string representing a computable quantum function on initial state; and a bit string that determines what measurement should be made on the resulting final state.
We call the second input a \emph{quantum program}.
Note that a quantum program, by our definition, is not quite a quantum algorithm.
A quantum algorithm is a recipe for transforming an input of any size; e.g., Shor's quantum factoring algorithm can be used to factor any number.
A quantum program, by contrast, acts only on inputs of a certain size.
In short, a quantum program classically encodes a unitary operator, and a quantum algorithm is a family of programs for different input sizes.

A commonly chosen universal set of quantum operations consists of arbitrary finite precision one qubit gates, along with the controlled-Z operation ($CZ$).

\begin{definition}\label{def:QAprimitives}
	Quantum Gates are strings that denote quantum operators that fall in either of the following two categories:
\begin{enumerate}
	\item The category of qubit rotations 
	\[
	R_j(\bm{\theta}) = \exp\left( -\frac{i}{\hbar} \bm{\theta}\cdot\bm{\sigma}_j \right) , 
	\]
	were $j$ indicates the qubit, $\bm{\theta} = (\theta_x,\theta_y,\theta_z)$, and $\bm{\sigma}_j = (X,Y,Z)$ is the vector of Pauli operators on qubit $j$.
	The size of this primitive in ``big Omega'' notation is $\Omega(m)$, where $m$ is the maximum precision of $\theta_x$, $\theta_y$, and $\theta_z$ in bits;
	\item the category of operators $CZ_{jk}$ of controlled-$Z$ operators between the $j^{\rm th}$ and the $k^{\rm th}$ logical data qubits. The size of this primitive is $\Omega(1)$.
\end{enumerate}
\end{definition}
Note that any gate amounts to a bit-string that records the type of operation, the qubit(s) it operates on, and if necessary the value of the angles in $\bm{\theta}$, up to a finite precision. 
The size function assumes that it is in general harder to implement a quantum unitary operation with higher precision, than one with lower precision. 
Hence, the higher precision of the gate, the higher its size value.
The two qubit gate, $CZ$, has a fixed constant size because its precision is fixed.

The second part of an inductive definition is the closure function. In other words, a method for constructing general set items from atomic ones.
\begin{definition}\label{def:Qcircuit}
	A Quantum Circuit can be constructed inductively as follows: 
\begin{enumerate}
	\item If $q$ is a gate  then it is a quantum circuit. 
	\item if $q_1$ and $q_2$ are quantum circuits, then their  composition, denoted by $q_2 \circ q_1$, is also a quantum circuit. 
\end{enumerate}
\end{definition}
A quantum circuit is therefore a bit-string that encodes (i.e., gives a method for implementing) a unitary operation.
Given a quantum circuit $q$ we denote the  unitary operator it encodes by $U_q$. It acts on a Hilbert space of size $2^{n}$ where $n$ is the size of the largest index in any gate in $q$ (recall that the index or indices of a gate establish which qubit(s) it acts on).

An important mathematical remark is that the set of quantum algorithms as defined above is \emph{freely generated}. This means that given a properly formed quantum circuit, there is only one way to construct it using the rules in definition \ref{def:Qcircuit}. This in turn implies that we can define a size measure as a recursive function on a quantum circuit, and that this value is always well-defined.

\begin{definition}\label{def:Qcomplexity}
	The size of a quantum circuit $q$, denoted by $C(q)$ or $|q|$ is a function from the set of all quantum circuits to the non-negative integers, and is defined as follows.
\begin{enumerate}
	\item If $q$ consists of a single gate, its size is the same as that of the gate (see Definition \ref{def:QAprimitives});
	\item if $q$ is a quantum circuit such that $q= q_2 \circ q_1$ then the size of this circuit is $|q_1| + |q_2|$.
\end{enumerate}
\end{definition}

In  Definition \ref{def:computableUnitary} below we describe computability and complexity of unitary operators in terms of the quantum programs that encode them.

\begin{definition}\label{def:computableUnitary}
A unitary operator $U$  is \emph{computable}, if there exists a quantum program $q$ such that $U_q = U$. 
Furthermore, the cost of $U$ is the size of the smallest quantum program $q$ such that $U_q = U$.
Finally, $U$ is $\epsilon$-approximable if there exists a program $q$ such that the fidelity $F(U, U_P) \geq 1 - \epsilon$. Similarly, the cost of approximating $U$ is  the size of the smallest program that approximates $U$.
\end{definition}

Using the concepts developed in this section we can easily and formally speak of quantum program size.
This is not a fully developed complexity theory, since it only speaks of the size of individual objects or programs. It is merely a useful tool to properly define quantum computers in the previous section. 
Also, we stress again that although we have used the language of circuits to discuss quantum programs, this is but one of many languages in which programs can be described. Regardless of which language we choose to represent a program, a quantum computer, as defined above, must be able to implement and run the program, possibly via a \emph{compiler}, i.e., a classical program that translates the the quantum program from our language to some internal representation appropriate for the hardware implementation. This translation might incur an additional cost, but it should always be a polynomially scaling penalty. This observation and restriction comes into play in our definition, where we state that the quantum computer must  implement the quantum program in a time that is polynomial in the size of the program.

\section{Criteria for building a quantum computer}\label{sec:ftnsc}

\noindent
One of the main aims of this paper is to establish the criteria that any implementation of a quantum computer must meet. 
The first set of these criteria were formulated by DiVincenzo, and mainly concern the circuit model of quantum computation, applied to qubits.
However, other models of quantum computation have been proposed since, and most likely new models will be formulated in the future. 
In this section we will discuss the set of criteria that any model of quantum computation must meet, now and in the future.
These are:
\setcounter{criterion}{-1}
\begin{criterion}
 Any quantum computer must have a quantum memory.
\end{criterion}
\begin{criterion}
 Any quantum computer must facilitate a controlled  evolution of the quantum memory, that allows for universal quantum computation.
\end{criterion}
\begin{criterion}
 Any quantum computer must include a method for cooling the quantum memory.
\end{criterion}
\begin{criterion}
 Any quantum computer must provide a readout mechanism for (non-empty) subsets of the quantum memory.
\end{criterion}\setcounter{criterion}{-1}
Criterion 0 establishes the \emph{conditio sine qua non} of any quantum computing device: its ability to dynamically represent a quantum state, in other words, have a quantum memory as defined in Definition \ref{def:qumemory}.
Criteria 1-3 establish further requirements imposed on this quantum memory.
These criteria are more general than DiVincenzo's criteria, and therefore more abstract. 
This means that we must give a metric that allows us to determine whether the criteria are satisfied. 
This metric requires two concepts, namely fault tolerance and scalability.

\subsection{Fault Tolerance}

\noindent
In practice, no device will ever be perfect. 
Random fluctuations induce errors in all aspects of the computation, including the quantum memory, the quantum evolution, and the readout.
If the device still operates in accordance with Definition \ref{def:kqubitQC} in the presence of errors, the device is called {\em fault tolerant}.
The size of the errors must typically be below a certain value in order to achieve fault tolerance, and this is called the fault tolerance {\em threshold}.
Fault-tolerance was first considered in classical computation, where fundamental results show that in the presence of faulty gates \cite{neumannX}, and even faulty gates and wires \cite{gacs}, classical computation can be done robustly.
The key to fault tolerance is to employ error correction effectively.
All properly functioning classical computers operate in a fault tolerant manner.

In quantum computing, due to the fragile nature of quantum information and entanglement, fault tolerance is much harder to achieve.
Indeed, many people initially believed that fault tolerant quantum computers are physically impossible \cite{unruh95}.
Nevertheless, it turned out that, as in classical computing, we can employ quantum error correction, and Shor and Steane where the first to show how to perform quantum error correction in principle \cite{Shor1, steane1, steane2}.
Each logical qubit is typically encoded in a number of physical qubits. 
A measurement of certain observables on the qubits then allows us to extract the {\em syndrome}, which tells us which, if any, error has occurred. 
For example, an error correction code can protect a logical qubit from a single error. 
However, the codes themselves consist of a number of physical qubits, all of which are prone to errors.
In order for quantum error correction to be beneficial, the probability of an error in the logical qubit must be smaller than the error probability on a single physical qubit, multiplied by all the distinct places the error can occur.
When this condition is satisfied, we can in principle concatenate the codes, which means that each physical qubit in the code is itself encoded, and so on.

However, in order to obtain fault tolerance, we also have to make sure that the quantum error correction code and the concatenation do not allow errors to multiply.
For example, when a qubit experiences a bit flip error, using that qubit as the control of a $CNOT$ will induce a bit flip error on the target qubit as well.
This means that we now have {\em two} errors, even though the $CNOT$ worked perfectly.
Quantum error correction codes must be designed such that error propagation and multiplication is kept under control.
Furthermore, all aspects of the quantum computer, including the memory, the evolution, and the readout, must be performed in a fault tolerant way.
All this must be achieved without compromising the polynomial scaling in the amount of resources required to perform the computation.
Some of the first results on fault-tolerance are due to Shor \cite{Shor2}, Kitaev \cite{kitaev97},  Steane \cite{steane3}, Gottesman \cite{gottesman}, and Preskill \cite{preskill98}.
How to achieve fault tolerance, and the numerical value of the threshold, depends on the paradigm and the type of error correction.
In Sec.\ \ref{sec:paradigms} we will discuss specific fault-tolerance thresholds for the particular paradigms.

\subsection{Scalability}

\noindent
The second essential characteristic of any quantum computer is {\em scalability}.
It is what allows us to move beyond the proof of principle experiment to a large-scale quantum computer that can solve interesting problems not within reach of classical computing.
While it is universally agreed that scalability is a desired, or even required, characteristic of potential quantum computing platforms, what exactly constitutes scalability is often subject to debate.
A textbook definition of scalability in computer science is that a system is scalable  if \emph{`its performance does not degrade significantly as [\ldots] the load on the system increases.'} \cite{menasce04}.
More generally, the cost of scaling a system to size $n$ can be described by a cost function. 
This function, and in particular its asymptotic growth rate, determines the degree of scalability of the system.

When discussing a quantum computing device using the above definition of scalability, one would say that the quantum computer is scalable if the resources needed to run the same polynomial-time algorithm on a larger input scaled polynomially on the the size of the input.
If the device, for reasons of having to cope with increased error-rates, decoherence, fundamental limitations on its size, etc., cannot compute the  algorithm using at most a polynomial overhead, we say that the device is not scalable.
The quantum information community often determines that it is the size of the quantum memory of a quantum computer that must be scaled.
Even without any further considerations, this also imposes a necessary scalability in the quantum control efficiency.
In particular, the more logical qubits are maintained in the device, the more parallel operations (or faster serial operations) are needed in order to keep the device fault-tolerant.
By this definition, there is currently no scalable quantum computing device.

While the above definition of scalability is widely accepted in the computer science and systems engineering communities, it is somewhat problematic when applied to quantum computing technology.
In particular, people often have a different definition of scalability in mind.
Considering that no scalable quantum computers exist yet, scalability usually refers to the future scalability of the proposed implementation of a quantum computer.
It is therefore a \emph{prediction}.

This issue was addressed recently by the Quantum Pontiff \cite{bacon09}, who presents an overview of several methods to analyze, evaluate and discuss the future scalability of a quantum computing device.
These include the economic cost of scaling these devices; the current knowledge of the technology used in the device; how this technology fits in the larger field of physics, chemistry, etc., and whether the technology has been used in scalable devices other than quantum computers.
These three forms of scalability can be understood as follows:
First, there is the {\em actual} scalability of a system. This is the scalability cost function of actually existing devices, according to the above definition.
Second, there is the {\em projected} scalability of a quantum computer architecture, which is given by the expected cost function.
As with the actual cost, this will be a function mapping the size of logical qubits to a resource cost.
It is similar in all other respects too, except that it refers to a hypothetical cost derived by projecting our advances in technology into the future.
In the absence of true quantum computers, the projected scalability is the quantity that is of most use to current designers of quantum computer architectures.
However, it is also the most difficult to quantify, due to inherent uncertainties in technological and economic predictions.
Finally, the {\em fundamental} scalability of an architecture is given by upper and lower bounds on the resource cost function that are direct consequences of the (known) laws of physics, in particular quantum mechanics.
These are absolute values, that upper and lower bounds can be derived formally from first principles, and generally the usefulness of the values for a particular device will rest in the size of the gap between the lower and upper bounds.

We can now state unambiguously when the criteria for quantum computing are met.
First, the device must be a fault tolerant quantum computer, which means that all the error probabilities for all the possible errors in a realistic error model are below the fault tolerance threshold. 
Consequently, the errors associated with each of the four criteria (memory, evolution, cooling, and readout) must be below this threshold.
Second, the device must be scalable. 
The specific type of scalability (actual, projected, or fundamental) depends on the objectives of the user.
For a prototype quantum computer one would require {\em actual} scalability up to a certain small number, while an industry interested in commercial production of quantum computers must meet {\em projected} scalability.

\subsection{The Criteria}

\noindent
We now discuss the four criteria that must be met when we want to build a scalable, fault tolerant quantum computer.

\begin{criterion}\label{crit:memory}
 Any quantum computer must have a quantum memory.
\end{criterion}

It is clear from Definition \ref{def:kqubitQC} that the quantum computer must be able to store the classical input bit string $\smash{s^{(n)}_{\rm in}}$ in such a way that it can be used in the computation. 
The storage device must be able to hold any computable quantum state that can be unitarily evolved from the input qubit string $\smash{|s^{(n)}_{\rm in}\ra}$.

Some states are known to have an efficient classical representation, such as the so-called stabilizer states \cite{gottesman97}. 
These can have large amounts of entanglement, and may be considered true quantum states.
Since we do not specify how the quantum memory operates, we may try to cheat the system by tracking only the efficient classical description of the state, and proceed with the computation using only these states.
However, if the memory holds only states of this type, the quantum computer operating with this memory cannot produce the output $\smash{p_P (s^{(m)}_{\rm out}|s^{(n)}_{\rm in})}$ efficiently.
This is a consequence of the Gottesman-Knill theorem \cite{gottesman97}.
A quantum computer requires the ability to store any computable quantum state, which in general, are not known to have efficient classical representations.

The quantum memory must be able to store any computable pure quantum state.
However, this does {\em not} mean that the quantum memory always {\em is} in a pure state.
The storage of a pure quantum state in an overall mixed system can be achieved in several ways. 
For example, a pure quantum state can be encoded in pseudo-pure states, or in decoherence-free subspaces.
Regardless of the encoding, the device generally needs to store a quantum state for prolonged periods of time.

\begin{criterion}\label{crit:evolution}
Any quantum computer must facilitate a controlled  evolution of the quantum memory, that allows for universal quantum computation.
\end{criterion}

According to Definition \ref{def:kqubitQC}, a quantum computer produces a probability distribution that depends on the unitary encoding $U_P$ of the program of interest $P$, and it must do so efficiently.
It is generally accepted that this implies that for many algorithms the same probability distribution cannot be simulated efficiently on a classical computer.
If this is indeed the case, then the evolution $U_P$ must be inherently quantum mechanical. 
Currently, we know of several ways to implement $U_P$, for example via a series of qubit gates, via adiabatic transformations, controlled measurements, etc. 
This process must act in accordance with the laws of quantum mechanics, and is ultimately controlled by the user or programmer of the device.
We therefore call this \emph{controlled quantum evolution}. 
This leads to Criterion \ref{crit:evolution}. 
The core distinguishing feature between various quantum computing paradigms is often the mechanism for performing the quantum evolution.
Other criteria (memory, cooling, and readout) are then chosen to support the specific implementation of the controlled evolution.
It does not matter whether the quantum evolution is described in the Schr\"odinger picture, the Heisenberg picture, or in an interaction picture.
These different pictures do not yield observable differences, and all lead to the same probability distribution $\smash{p_P (s^{(m)}_{\rm out}|s^{(n)}_{\rm in})}$.

At this point, we should note that any type of controlled evolution implies that the quantum computer must incorporate a clock of some sort.
Moreover, as quantum computing is currently understood, each part of the computer must be synchronized to the same clock.
It is, however, possible to implement distributed quantum computing between parties that move relativistically with respect to each other by encoding the quantum information in Lorentz invariant subspaces \cite{kok05}.
A general discussion on the role of reference frames in quantum information can be found in 
\cite{bartlett07}.

\begin{criterion}\label{crit:cooling}
 Any quantum computer must include a method for cooling the quantum memory.
\end{criterion}

Once we have built a quantum computer, we most likely want to use it more than once. 
Since any real quantum computer has a finite size $k$, the preparation of a computation therefore includes the erasure of the previous computation.
The entropy generated in this procedure must be extracted from the quantum computer, and we call this (information theoretic) cooling.
In addition to the entropy generated by the erasure of previous computations, entropy may leak into the quantum memory via unwanted and uncontrolled interactions with the environment. 
Such entropy leaks cause errors in the computation, which may be removed using quantum error correction procedures.
This is also a form of cooling. 
Any real quantum computer must therefore satisfy Criterion \ref{crit:cooling}.

The concept of cooling encompasses both the initialization of the quantum memory and error correction during the computation.
Furthermore, the boundary between initialization and error correction is fuzzy.
For example, in the one-way model of quantum computation, it can be argued that the initialization process is the construction of a high-fidelity graph state. 
In this case the initialization may include the preparation of the qubits in a ground state, as well as the entangling interactions and the various entanglement distillation procedures that are used to make the graph state.
Alternatively, one may argue that initialization means only the preparation of the qubits in the pure ground state $\ket{0}$. The creation of the graph state then falls in the category of controlled quantum evolution, augmented by quantum error correction.
Whether it is more natural to view the entire graph state creation as the initialization procedure, or only qubit initialization depends often on the physical implementation.
In optical lattices it may be more appropriate to interpret initialization as the creation of the cluster state, whereas systems in which each two-qubit interaction must be invoked separately favor qubit initialization as the natural interpretation.

A consequence of this fuzziness between initialization and error correction is that it is not important what pure state the system is cooled to.
In practice, this will depend on the most accessible states of the physical implementation.
Once the system is cooled down sufficiently, the different types of control (both quantum evolution and further error correction) can bring the quantum memory into any desired state.

\begin{criterion}\label{crit:readout}
 Any quantum computer must provide a readout mechanism for subsets of the quantum memory.
\end{criterion}

According to Definition \ref{def:kqubitQC}, a quantum computer produces a classical bit string $\smash{s_{\rm out}^{(m)}}$ as the output. 
This implies that the quantum computer includes a mechanism that translates the output state of the quantum memory to a classical bit string. 
This is done via measurement, and in the context of computing we call this readout.
Therefore, Criterion \ref{crit:readout} must be satisfied for any quantum computer.

Similar to cooling, what is considered readout is a rather fluid concept.
Much of error correction involves the measurement of large parts of the quantum memory, but the end user is generally not interested in these measurement outcomes.
We therefore may regard this type of readout as part of the cooling mechanism.
Again, the most natural interpretation depends on the physical implementation of the quantum computer.

Since the different paradigms may differ in the type of control they require, they may also differ in the required readout abilities.
In general, the more restrictive the controlled quantum evolution requirement is, the less restrictive the readout requirement needs to be.
For example, the ability to do single-qubit measurements on an arbitrary basis can be exchanged with the ability to do single-qubit measurements on a predefined axis, if one has also the ability to do arbitrary single qubit rotations.

\section{Computational Paradigms}\label{sec:paradigms}

\noindent
There are potentially many ways to implement a quantum computer. 
For example, we can use trapped ions with optical control fields, photons in linear optical networks, electrically controlled quantum dots, etc.
The specific way the quantum computer is constructed is called the {\em architecture}.
It encompasses all the details of the implementation.
Apart from the architecture, there is another useful distinction in different types of quantum computers.
Not all types of quantum computers treat the quantum information in the same way, and we call the different ways of implementing the computation at the abstract computational level {\em paradigms}.
The paradigm can be interpreted as the way the computation $\smash{s_{\rm in}^{(n)} \to s_{\rm out}^{(m)}}$ is decomposed into primitive elements.
As an example, in Definitions \ref{def:QAprimitives} and \ref{def:Qcircuit}, the primitives are chosen as the single-qubit rotations, supplemented by the two-qubit $CZ$ gate.
These can be translated into other primitives, associated with the different paradigms.

Finally, there is what we refer to as \emph{data abstraction and encoding}.
Error correction schemes fall into this category. 
For example, CSS, stabilizer, and topological codes are all different ways to encode data.
These encodings can be used with various (though not necessarily all) paradigms. 
A particular case worth mentioning is topological quantum computing, which was first introduced as the coupling of the topological data encoding with the anyonic architecture. 
Though first described together, it has been shown that topological codes can be used in other architectures and paradigms, such as the one-way model of quantum computing, using photons and matter qubits instead of anyons.

The examples of different paradigms we will discuss shortly are the circuit model, globally controlled quantum computing, measurement-based quantum computing, and adiabatic quantum computing.
Each paradigm must meet the criteria, but they are typically met in slightly different ways.
Furthermore, how a particular device satisfies the criteria determines the most natural quantum computing paradigm.
In this section we will discuss the various paradigms of quantum computing, and how the criteria for quantum computing must be met.

\subsection{The circuit model}\label{sec:circuit-model}

\noindent
In the circuit model, the computation is decomposed into logical gates that are applied successively to the qubits. 
This is commonly represented graphically as a circuit, where each horizontal line denotes the time evolution of a qubit, and logical gates are symbols on the lines.
The circuit model is arguably the most natural way to visualize a quantum computation, and the universality proof of other paradigms often proceeds by reduction to the circuit model.

The key to a circuit-model description of quantum computing is a series of results, showing that any unitary operator on $n$ qubits can be decomposed into a series of arbitrary single-qubit rotations, and two-qubit entangling gates \cite{barenco95,barenco95b,divincenzo95}.
Such a restricted set of gates is called {\em universal}.
In addition, it was shown that almost any two-qubit entangling gate can be used to construct a universal set of gates \cite{Deutsch95,lloyd95}.
Typical choices of the entangling gate are the $CNOT$ and the $CZ$.
This universal set of operations is infinite, since it contains all possible single qubit rotations on any one qubit.
The Solovay-Kitaev theorem states that for any unitary operation $U$ there exists a finite set of gates that can efficiently implement $U$ to arbitrary precision \cite{kitaev97}.
The most general form of this theorem was proved in Appendix 3 of \cite{nielsen00}, and for a history of the theorem see \cite{dawson06}.
The implication of the theorem is that we can construct a quantum computer based on a finite set of gates.

In order to achieve a complete computation device it is necessary to supplement the universal set of gates mentioned above with two more primitives, namely measurements and cooling.
It is often assumed that the measurements are {\em von Neumann} measurements, which leave the system in the eigenstate corresponding to the measurement outcome.
This type of measurement can act as a cooling mechanism as well, since it transforms mixed states into pure states.
In general, however, von Neumann measurements may be difficult to implement (for example when the quantum information is carried by a photon, which is usually destroyed by  the detector), and we will assume here that measurement and cooling are independent requirements.

Fault tolerance thresholds in the circuit model (as well as in other paradigms) depend very much on the implementation of the quantum error correction codes. 
Consequently, the thresholds vary substantially. 
Early calculations yielded thresholds for the error probability per gate around $10^{-6}$ \cite{kitaev97,preskill98,knill98,aharonov99}, and $10^{-4}$ \cite{gottesman97}. Steane
 proved a threshold of $3\cdot 10^{-3}$ \cite{steane03}, and Knill derived a threshold of about one percent \cite{knill04b}. 
Models that allow only nearest-neighbour interactions have thresholds on the order of $10^{-4}$ to $10^{-5}$ \cite{svore:022317}.

\begin{table}[t]
\caption{\label{tab:circuit} Quantum computing criteria for the circuit model.\smallskip}
\begin{indented}
\item[]\begin{tabular}{@{}ll}\br
{\bf C0.} & Identifiable, and addressable, individual qubits; \\
{\bf C1.} & the ability to implement a universal set of gates; \\
{\bf C2.} & the ability to reduce the entropy of the state of any qubit; \\
{\bf C3.} & the ability to measure any single qubit in the computational basis. \\
\br
\end{tabular}
\end{indented}
\end{table}

Summarizing, the requirements for universal fault-tolerant quantum computation using the circuit model are given in Table~\ref{tab:circuit}.
We have put `qubits' in quotes, since there are circuit-model proposals that act on qudits or continuous variables \cite{lloyd99}. 
However, these proposals can always be recast in terms of qubits, including continuous variables (with finite precision).
For qubits, this list reduces to the DiVincenzo criteria, where our Criterion 0 encompasses DiVincenzo's Criteria (i) and (iii); our Criterion 1 is equal to Criterion (iv); our Criterion 2 is equivalent to Criterion (ii), and our Criterion 3 is DiVincenzo's Criterion (v).

\subsection{Global control}\label{sec:global-control}

\noindent
The second paradigm we consider here is usually called `globally-controlled quantum computing', or `global control', for short.
In 1993 Seth Lloyd described the fundamental idea behind this paradigm, which differs significantly from the circuit model in that it does not employ a universal set of single- and two-qubit gates \cite{lloyd93}.
Instead, the quantum memory consists of units called {\em quantum cells}, which have some controllable (often nearest-neighbour) interactions that can be switched on and off.
The quantum evolution associated with the algorithm then proceeds via operations that act on all quantum cells indiscriminately.
As an example, consider a one-dimensional spin-$\frac{1}{2}$ lattice (this can be a crystal, optical lattice, etc.), often called a {\em spin chain}.
Suppose further that the method of addressing the spins in the lattice is through some magnetic field.
A homogenous field affecting all spins identically is typically easier to achieve than a gradient field that couples to only a few, or even a single lattice spin.
Similarly, the distance between interacting spins is much shorter than optical wavelengths, making individual optical control of the spins extremely challenging.
A global-control paradigm offers clear practical benefits in these situations.

In Lloyd's original scheme, the spins in the one-dimensional lattice are of three different types, or species, $A$, $B$, and, $C$, arranged in cyclical repeating order with switchable nearest-neighbour interactions.
Each species can be addressed independent from the other species. 
A setup that accommodates this type of control is a crystal or polymer with three different species of nuclei, and where each nucleus is coupled to its neighbours.
The controlling mechanism then consists of magnetic pulses tuned to the resonance frequencies of the target nuclei.
Lloyd showed that even such limited control over a homogenous system like this allows for universal quantum computation.
The scheme involves initializing the system to a particular state where the entire lattice, except for a small region, is set to a fiduciary initial state.
The small region is set to a state known as the \emph{control pointer}.
Logical operations then consist of applying homogenous pulses that act only non-trivially in the vicinity of the control pointer state.
These operations can change the state of logical qubits in the vicinity of the pointer, or move the pointer up and down the chain.
It was further shown by Benjamin \cite{Benjamin:2004eu,Benjamin:2004wd,Benjamin:2004nx,Benjamin:2003oq} that two different spin species, $A$ and $B$, still allows for universal quantum computation.
Other possible implementations are discussed in \cite{Vollbrecht:2006lq,cp07,fitzsimmons07,Fitzsimons:2006ul}.

Readout is done through a global, or \emph{bulk}, measurement on a cell species.
This means that the state of any one particular cell is not readily available, but rather it is possible to make a measurement of a global character that does not distinguish between cells.
An example measurement of this type is the bulk magnetization of a spin ensemble. 
Mathematically, it is similar to projecting onto eigenspaces spanned by computational basis states of a certain Hamming weight.
This might not seem powerful enough to give the output of a computation, but it is indeed possible to use the finite nature of the cell chain, and the particular boundary conditions, to extract the result of the computation \cite{lloyd93,Benjamin:2004eu}.
Another possible technique includes the use of spin-amplification \cite{perez06}.
Quantum Cellular Automata (QCA) are similar to spin chains, with the added constraint that the evolution is not only homogenous in space, but also in time.
In other words, the global operation is a repeating cycle of a set of operations.
QCA are an important theoretical construct, in that they are the natural generalization of the classical model of computation based on cellular automata \cite{cp07}.

Quantum error correction in globally controlled systems was discussed by Bririd {\em et al}.\ \cite{bririd}.  
Fitzsimmons \cite{fitzsimmons07,Fitzsimons:2006ul} proved the existence of a threshold, and Kay \cite{kay-2007,kay-2005} proved a fault tolerance theorem with a threshold of $10^{-11}$.
Both fault tolerant protocols require the ability to cool spins using global control. 
One way to achieve cooling is to endow all spins (or at least a large subset of them) with a a third, unstable state $\ket{2}$.
Spontaneous emission of a photon from this state then produces the required ground state $\ket{0}$. 
It is not normally assumed that these emissions can be detected.
However, this does not affect the initialization procedure.
QCA have been shown to be universal for quantum computing \cite{Vollbrecht:2006lq,watrous,cp07}, but it remains an open question whether they can be implemented in a fault tolerant manner.
The requirements for universal fault-tolerant quantum computation using the globally controlled array model are summarized in Table~\ref{tab:global}.

\begin{table}[t]
\caption{\label{tab:global} Quantum computing criteria for global control models.\smallskip}
\begin{indented}
\item[]\begin{tabular}{@{}ll}\br
{\bf C0.} & Identifiable and {\em globally} addressable individual quantum cells; \\
{\bf C1.} & the ability to implement a universal set of global operators; \\
{\bf C2.} & the ability to reduce the entropy of the global state of a species of cells;\\
{\bf C3.} & the ability to make a global measurement on a species of cells. \\
\br
\end{tabular}
\end{indented}
\end{table}

\subsection{Measurement-based quantum computing}\label{sec:measurement-based}

\noindent
Measurement-based quantum computing has its origins in at least two converging lines of research. 
First, it was realized in the quantum computing community that two-qubit gates are generally far more difficult to implement with high fidelity than single-qubit gates. 
This led to the concept of gate teleportation \cite{gottesman99}, in which the quantum channel of a teleportation event is modified to induce a specific gate on the teleported qubits. 
The most famous application of gate teleportation is the demonstration that it is possible to build a quantum computer with only single photons, linear optical elements, and photon counting \cite{knill01}. 
The second line of research that led to measurement-based quantum computing was the study of practical applications of large-scale entanglement with a lattice structure \cite{briegel01}. 
These so-called {cluster states} appear naturally in optical lattices, and many other regular structures that are characterized by nearest-neighbour interactions (such as the Ising interaction).

Measurement-based quantum computation relies on the preparation of a large entangled state, the cluster state, which is typically a regular lattice where each vertex represents a qubit initially in the state $(\ket{0}+\ket{1})/\sqrt{2}$, and each edge a $CZ$ operation \cite{raussendorf01}.
The computation proceeds by making single-qubit measurements on a subset of the qubits.
The measurement outcomes determine the single-qubit observables for the next measurement round, and so on.
Many classes of large entangled states have been identified as universal resources for quantum computing.
The quantum program is encoded entirely in the single-qubit measurement bases \cite{raussendorf03,hein04}.
Because of this measurement-driven approach, and the fundamental irreversibility of the measurement procedure, this is also called the {\em one-way model} of quantum computing. 
One of the major advantages of the one-way model over the circuit model is that the universal resource state, i.e., the cluster state, does not carry any information about the computation.
It is therefore possible to create these states by any (efficient) means, not necessarily using high-fidelity deterministic two-qubit gates \cite{nielsen04,browne05,barrett05}.
This can significantly reduce the requirements for building a quantum computer.

The fulfillment of the four criteria for building a quantum computer in the one-way model is somewhat fluid, in the sense that we can interpret certain aspects of the model as falling under different criteria.
First, the quantum memory of a quantum computer based on the one-way model consists of qubits, although the model can also be defined on qudits \cite{zhou03,tame06} and qunats \cite{menicucci06}.
The qubits must be addressable to the extent that any single-qubit measurement in the equatorial plane of the Bloch sphere, as well as a computational basis measurement, can be reliably performed.
Similar requirements exist for measurement-based protocols that are based on other types of information carriers, such as continuous variables \cite{menicucci06}.
Second, the controlled quantum evolution is somewhat hidden in measurement-based quantum computation.
We start with a large entangled resource state that is typically a stabilizer state, which permit an efficient classical description. 
The measurements will remove qubits from the state, and in doing so drive the entangled resource state to different states that are typically no longer efficiently describable as stabilizer states.
Another sense in which measurement-based quantum computing needs controlled quantum evolution is in the creation of the entangled universal resource.
If we use probabilistic gates, we have to put up with an inherent lack of control.
However, we can choose efficient strategies that allow us to induce the control necessary to create the universal resource \cite{gross06}.
Alternatively, we can use near-deterministic gates with additional purification and distillation \cite{lim06}.
Third, the creation of the entangled resource can also be regarded as cooling to the ground state of a suitable many-body Hamiltonian.
Still, even after the cluster state has been produced we have to allow for quantum error correction protocols, since the cluster state must be protected from errors while its qubits are waiting to be measured.
Finally, the readout mechanism is of course central to the measurement-based model.
Not only does the one-way quantum computer need the readout for the final step of retrieving the outcome of the computation, the computation itself is driven by the measurements.
The criteria are summarized as in Table~\ref{tab:mbqc}.
Here we encounter a rare occasion of a paradigm for which one criterion implies another.
In this case, Criterion 1 implies Criterion 3. 
This indicates that measurement-based quantum computation may in some cases be easier to implement than other paradigms.

\begin{table}[t]
\caption{\label{tab:mbqc} Criteria for measurement-based quantum computing.\smallskip}
\begin{indented}
\item[]\begin{tabular}{@{}ll}\br
{\bf C0.} & Identifiable, and addressable, individual qubits; \\
{\bf C1.} & the ability to implement single-qubit measurements in a large subset of bases; \\
{\bf C2.} & the ability to cool the quantum memory to a universal entangled resource state; \\
{\bf C3.} & the ability to measure any single qubit in the computational basis. \\
\br
\end{tabular}
\end{indented}
\end{table}

The one-way model of quantum computing can be made fault tolerant as well. Two threshold theorems by Nielsen and Dawson (2005) give the maximum allowable errors when the cluster states are created with noisy deterministic linear optical entangling gates, and when only probabilistic (noisy) linear optical entangling gates are available \cite{nielsen05}. 
Raussendorf and Harrington derived a threshold theorem for general two-dimensional cluster states, and found a maximum error of 0.75\% \cite{raussendorf06,raussendorf07,raussendorf07b}.


\subsection{Adiabatic control}\label{sec:adiabatic}

\noindent
The final paradigm  we consider here is adiabatic quantum computing.
The main difference from the previous paradigms is that in the adiabatic model the quantum information is not processed in discrete time steps (i.e., gates), but in a continuous fashion. 
Of course, all gates must also operate continuously in the temporal domain in any practical implementation, but the important distinction here is that in non-adiabatic paradigms the quantum program is defined \emph{procedurally}.
In other words, an algorithm is composed of discrete time steps, at which specific operations are carried out.
Adiabatic quantum computing is a complete departure from this line of thinking.
The core of this paradigm is the {\em adiabatic theorem}, first developed and proven by  Born and Fock \cite{born}. 
The theorem states that a physical system will remain in its instantaneous eigenstate if any given perturbation  acting on it  does so slowly enough with respect to the gap between the minimal eigenvalue and the rest of the Hamiltonian's spectrum.

The basic idea of the adiabatic paradigm is that instead of carrying out an algorithm that takes you to the desired output via gates, one has to give a Hamiltonian $H_f$ whose ground state represents the solution to the computational problem.
The quantum computer starts in the ground state of a Hamiltonian $H_0$, and the computation proceeds by adiabatically changing the initial Hamiltonian $H_0$ into $H_f$.
The paradigm has a very distinct elegance, since we do not have to state the procedure, but rather the desired output in terms of a Hamiltonian.
In classical computing this approach leads to various benefits, for example, easily provable correctness of algorithms.

The standard example of this paradigm is the adiabatic implementation of a Grover search.
The problem is to find a marked bit string $s$ of size $n$, out of a possible $N=2^n$ bit strings.
Grover showed by explicit construction that this can be solved in time $O(\sqrt{N})$, whereas the most efficient classical algorithm scales as $O(N)$.
In the adiabatic paradigm we choose $H_f = \unity - \ket{s}\bra{s}$.
The ground state of this Hamiltonian is $\ket{s}$, i.e., the solution to the problem.
Moreover, the Hamiltonian can be constructed without explicit knowledge of $\ket{s}$ (which would defeat the purpose), but rather by properly encoding the condition that marks $s$ as the solution.
Since there is no \emph{a priori} knowledge of which string is the marked one, we start the system in a superposition $\ket{\psi_0}$ of all possible strings,
and we let $H_0 = \unity - \ket{\psi_0}\bra{\psi_0}$.
Evolving adiabatically from $H_0$ to $H_f$, and measuring the final state will provide the desired outcome $s$.
Roland and Cerf calculated the time constraints needed in order to maintain adiabaticity, and showed that total running time of the algorithm is $O(\sqrt{N})$ \cite{cerf02}.
It is straightforward to adapt this algorithm to solve any $NP$ problem, again in $O(\sqrt{N})$ time (where $N$ is the size of the search space, or set of possible solutions).
Recently, Aharonov demonstrated a procedure for adapting any quantum circuit algorithm into an adiabatic procedure \cite{aharonov07}.
However, the straightforward translation of algorithms can produce unwieldy results.
Even though all paradigms have the same computational power, some algorithms are better described in one model rather than another. 
In the case of adiabatic quantum computation, the natural algorithm is Grover's search.

From an implementation standpoint, perhaps the biggest drawback of adiabatic quantum computation is the very specific, and non-trivial, requirements on $H_f$.
While in the quantum circuit model all quantum computation can be reduced to single- and two-qubit gates, in the adiabatic model we often need many-body interactions.
For example, three-body interactions are required for the naive implementation of satisfiability, and up to five-body interactions are necessary for general quantum circuit simulation.
If we implement a fault tolerant adiabatic computation, the many-body interactions may be even higher.

The error correction schemes that are developed for adiabatic quantum computing are so far based on using quantum error detecting codes in the Hamiltonian, ensuring that any logical error would drive the energy of the system considerably upward \cite{jordan06}.
This should ensure that the system, assuming it stays sufficiently cool, does not leave the ground state.
It is in principle possible to use longer and longer codes, which would protect against larger and larger entropy in the system.
However, this also requires higher many-body terms in the Hamiltonian.
Progress is being made towards complete fault tolerance for adiabatic quantum computing \cite{lidar08}, but there currently exists no threshold theorem.
We summarize the criteria for universal fault-tolerant adiabatic quantum computation in Table~\ref{tab:adiabatic}.
However, since the precise requirements for fault tolerance are not known at this time,
these criteria may change.
In particular, the required size of the many-body interaction is not known at this time.

\begin{table}[t]
\caption{\label{tab:adiabatic} Criteria for adiabatic quantum computing.\smallskip}
\begin{indented}
\item[]\begin{tabular}{@{}ll}\br
{\bf C0.} & Identifiable, and addressable, individual qubits; \\
{\bf C1.} & the ability to implement arbitrary Hamiltonians consisting of many-body interactions \\
 & terms; \\
{\bf C2.} & the ability reduce the entropy of the quantum memory to the ground state of a \\ 
& prescribed Hamiltonian; \\
{\bf C3.} & the ability to measure any single qubit in the computational basis. \\
\br
\end{tabular}
\end{indented}
\end{table}

\subsection{Hybrid architectures}\label{sec:hybrid}

\noindent
While many architectures for quantum computers use only one particular paradigm, this is not necessary in general.
It is possible that specific implementations of a quantum computer 
use a combination of paradigms for different parts of the computation.
Several hybrid schemes have been proposed already, and we mention a few of them here.

First, different paradigms may be used to implement a quantum computer, and to connect the several computers in a network.
For example, a circuit model-QCA hybrid architecture was proposed by Laflamme and Cory for a universal quantum computer developed in the near future. 
Such a device would have several universal quantum registers, each with the same number of  logical qubits.
These registers are linked to each other via spin chains, which are in turn controlled as a QCA, and act as conveyer belts for information between the universal registers.

Second, different paradigms may be used for creating the universal quantum evolution on the one hand, and the error correction protocols on the other.
For example, a hybrid architecture using the one-way model and the circuit model was proposed by Campbell and Benjamin \cite{campbell08}. 
This is based on a distributed scheme where distant qubits are entangled via optical path erasure, but instead of a single qubit at each distant site there may be several qubits.
These few-qubit processors process quantum information using the circuit model.
An earlier version of such an architecture is the broker-client scheme for efficiently creating large cluster states \cite{benjamin06a}.
The number of qubits per site may in principle become quite large, allowing error correction and other low-level information management at the site level, and having only logical or higher level operations occur via optically entangling remote operations.

Third, we can even use different paradigms in several aspects of the quantum evolution.
For example, we can decompose the evolution in terms of a quantum circuit, but use adiabatic control to implement the universal set of quantum gates \cite{gauger08}.
In short, even if none of the existing paradigms suit a particular physical setup perfectly, it might still be possible to tailor a hybrid scheme.
Finally, we should note that there are quite possibly many more universal quantum computational paradigms.

\subsection{Quantum communication}

\noindent
The four criteria for quantum computing also apply to quantum communication, albeit in slightly modified form. 
First, we can regard quantum communication as a form of distributed quantum computing, even though communication protocols are typically much simpler than general quantum algorithms.
We therefore consider the criteria explicitly for the case of distributed quantum computing. 
Clearly, the criteria are then {sufficient} for quantum communication. 
However, they are not quite {necessary}.
Quantum communication requires quantum memories, cooling, and readout, but only a restricted set of quantum algorithms need to be performed.
We therefore modify Criterion \ref{crit:evolution} to 

\medskip
\noindent
{\bf Criterion 1a.} Any quantum communication device must facilitate a restricted set of controlled quantum evolution of the quantum memory.
\medskip

\noindent
The restricted set consists of the identity, and simple operations, such as the swap operation.

The measure that determines whether the criteria are satisfied are still scalability and fault tolerance.
Scalability is almost immediate: when Alice and Bob are able to communicate a single qubit at a cost $C$, they will be able to communicate $N$ qubits at cost $NC$, simply by repeating the procedure for communicating a single qubit.
Fault tolerance for quantum communication can also be defined.
In particular, this is illustrated by the security proof of BB84 by Shor and Preskill \cite{shor00}.
Security is defined in terms of the maximum allowable mutual information between Alice and Bob on the one hand, and the environment on the other.
Error correction can then be used to minimize the mutual information.
Shor and Preskill found a maximum allowed error of about 11\%.
This can be regarded a threshold theorem for BB84.

\section{Selecting a Paradigm}

\begin{figure}[t]
  \begin{center}
  \begin{psfrags}
      \psfrag{a}{\small\sf Start}
      \psfrag{b}{\small\sf Scalability}      
      \psfrag{c}{\small\sf Addressability}      
      \psfrag{d}{\small\sf Control}      
      \psfrag{e}{$\!\!$\footnotesize\sf Circuit Model}      
      \psfrag{f}{\footnotesize\sf One-way QC}      
      \psfrag{g}{$\!\!$\footnotesize\sf Global Control}      
      \psfrag{h}{\footnotesize\sf Adiabatic QC}      
      \psfrag{i}{\small\sf Monolithic}      
      \psfrag{j}{\small\sf Local}      
      \psfrag{k}{\small\sf Non-adiabatic}      
      \psfrag{l}{\small\sf Modular}      
      \psfrag{m}{\small\sf Global}      
      \psfrag{n}{\small\sf Adiabatic}      
       \epsfig{file=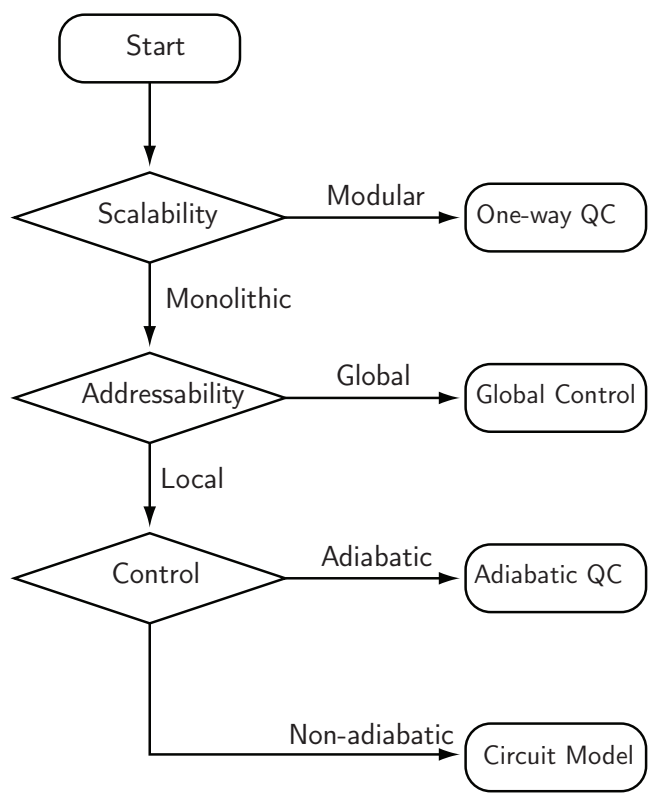,width=8.5cm}
  \end{psfrags}
  \end{center}
 \caption{Example of a decision tree to determine which quantum computing paradigms is most suitable for a give physical setup.} 
 \label{fig:flowchart}
\end{figure}

\noindent
With the availability of various paradigms for quantum computing, the natural question for any experimentalist is `which paradigm is best suited for my physical system'.
In order to help answer this question, we present a decision tree (shown in Fig.~\ref{fig:flowchart}) that may act as a rough guide towards implementing fault tolerant quantum computation.
To this end, we discuss some aspects of scalability, addressability of the qubits, and the type of qubit control that is most natural for the physical system.

\subsection{Monolithic vs. Modular Scalability}

\noindent
Scalability is required for any implementation of a quantum computer.
Here, we consider two types of scalability, namely {\em monolithic} scalability, and {\em modular} scalability.

Scalability is monolithic when the quantum computer is a single device, and we can increase the size of the device while still satisfying the criteria.
One example of a monolithically scalable setup is solid-state NMR.
Here, each nucleus is a physical qubit, and scaling the device simply implies using a larger solid with more nuclei. 
Other examples include optical lattices, quantum dots, and superconducting qubits.
An example of a system that is not monolithically scalable is an atom in a cavity.
We cannot double the number of atoms in the cavity without altering the physics of the device.
However, atoms in cavities can exhibit modular scalability.
We can increase the number of cavities (with atoms inside), which in turn can be entangled remotely, for example using mediating photons.
A scalable quantum computer can then be built by using various cavities, or more abstractly and in general, various {\em modules}.
There is a certain amount of arbitrariness in determining whether a system is modular or monolithic.
For example, a full-scale quantum internet can be interpreted as a modular system, even if each module is a monolithic $k$-bit quantum computer.
Whether a system should be regarded modular or monolithic therefore depends in a large part on the context.

While it is possible to force both monolithic and modular quantum devices into any of the paradigms above, some paradigms can be seen as more natural.
A physical setup that exhibits modular scalability is best suited for the measurement-based quantum computing paradigm of Sec.\ \ref{sec:measurement-based}. 
A setup that is monolithically scalable may be better suited for the circuit model (Sec.\ \ref{sec:circuit-model}), or global-control (Sec.\ \ref{sec:global-control}), depending on the other device properties.
Setups that are somewhere between monolithically and modularly scalable can benefit from a hybrid approach (see Sec.\ \ref{sec:hybrid}).

\subsection{Addressability}

\noindent
The quantum memory of the device, whether it be monolithic or modular, must be addressable in some form.
We have seen that different paradigms place different requirements on the precise character of this addressability.
In the circuit model and the measurement-based model, each qubit must be addressed individually, while in the global control model the quantum memory is addressed without discriminating between qubits.
This typically places restrictions on the type of interactions between the qubits (see Sec.\ \ref{sec:global-control}).

It is a natural assumption that each module in a modular device can be distinguished and addressed independently from each other module.
Hence, addressability is really only a concern for the physical qubits \emph{within} each module, and is of greatest concern for monolithic setups.
Therefore, the paradigm best suited for monolithic quantum systems with limited addressability is globally controlled quantum computing.

Another important issue is the level of addressability.
Like other features discussed before, addressability is not a binary condition, but rather there is a full gradient of possibilities.
On the one extreme, there is complete individual addresability of each individual subsystem, qubit \emph{etc.}
On the other extreme, each subsystem is completely indistinguishable from any other.
In between, we have systems, and architectures, where there are two, three, or more \emph{distinguishable} species. 
Furthermore, these may be spatially ordered in a homogenous or non-homogenous fashion. 
Physical system examples of these differences could be a carbon nano-tube, where the individual carbon nuclei are the fundamental quantum subsystems, each indistinguishable from each other; and a polymer consisting of two or three different nuclei arranged in a repeating pattern.
Various proposals, as discussed in Sec \ref{sec:global-control}, are available, each more or less relevant depending on the exact nature of the physical device.

\subsection{Quantum Evolution Implementation}

\noindent
Finally, we consider the type of control that we may have over a device.
This depends on various aspects of the architecture.
For instance, modular devices, and those with limited addressability, impose restrictions on the type of control available to the user or programmer of the quantum device.
Even without these constraints, however, it is possible to choose between two fairly distinct methods of quantum control, namely adiabatic or non-adiabatic quantum evolution.
In our decision tree in Fig.~\ref{fig:flowchart}, this choice is the final bifurcation that indicates which quantum paradigm is most suitable for a particular device.
Again, the distinction between adiabatic and non-adiabatic control can be blurry in some cases. 
As an example of a hybrid approach to quantum control we refer to Gauger {\em et al}., who propose an implementation in which the computational paradigm is the circuit model, but (some of) the individual gates are operated in an adiabatic manner \cite{gauger08}.

In general, the decision tree should be taken at best as a help to guide experimentalists towards the most suitable implementation given their experimental setup.
As new paradigms are developed, the decision tree will grow more branches, and other questions about the capabilities of the physical devices must be answered.
Nevertheless, the four criteria themselves are independent of the computational paradigm, and should therefore once more be regarded the central dogma for implementing a quantum computer.

\section{Conclusions}

\noindent
The DiVincenzo criteria have been extremely influential in focussing the theoretical and experimental research in quantum computing. 
However, since the initial formulation of the criteria for the circuit model, several new paradigms for quantum computation have been invented or developed further. 
As a consequence, the original criteria are sometimes violated in certain paradigms. 
In this paper, we have generalized the DiVincenzo criteria to take into account new paradigms, such as the one-way model, globally controlled quantum computing, and adiabatic quantum computing. 
We distilled the criteria down to four general requirements, namely the availability of a quantum memory, the ability to induce a (near) unitary evolution, the ability to implement (information-theoretic) cooling, and readout of the quantum memory. 
These criteria are derived directly from a new definition of a quantum computer. 
We distinguish between an ideal quantum computer, which has arbitrarily large size, and a $k$-bit quantum computer that can have a physical implementation. 
The desiderata that determine whether the criteria are met are fault tolerance and scalability.

In addition to the four criteria for quantum computing, we constructed a decision tree that may help experimentalists decide which paradigm is the most natural for a particular physical implementation. 
This decision tree will have to be updated whenever new paradigms for quantum computing are invented. 
However, the criteria are independent of new paradigms.

\section*{Acknowledgments}

\noindent
The authors wish to acknowledge valuable discussions with and comments from Sean Barrett, Niel de Beaudrap, Earl Campbell, Donnie Chuang, Irene D'Amico, David Deutsch, Richard Jozsa, Andrew Steane, Terry Rudolph, Stefan Weigert, and David Whittaker.

\section*{References}
\bibliographystyle{plain}
\bibliography{criteria}

\end{document}